\documentclass[pre,twocolumn,amsmath,amssymb,showpacs]{revtex4}
\usepackage{fancyhdr}
\usepackage{graphicx}
\usepackage{amsmath}
\usepackage{bm}
\usepackage{subfigure}
\usepackage{lastpage}
\newcommand{\bra}[1]{\langle #1|}
\newcommand{\ket}[1]{|#1\rangle}

\begin{document}

\title{Light scattering from three-level systems: The T-matrix of a point-dipole with gain}

\author{Tom \surname{Savels}}
\email{t.savels@utwente.nl}
\author{Allard P. \surname{Mosk}}
\author{Ad \surname{Lagendijk}}
\affiliation{\small{Complex Photonic Systems, Dept. Science and
Technology\\\small University of Twente, PO Box 217, 7500 AE
Enschede, The Netherlands.}}

\date{January 25th, 2005}

\begin{abstract}
We present an extension of the T-matrix approach to scattering of
light by a three-level system, using a description based on a
Master equation. More particularly, we apply our formalism to
calculate the T-matrix of a pumped three-level atom, providing an
exact and analytical expression describing the influence of a pump
on the light scattering properties of an atomic three-level
system.
\end{abstract}

\maketitle \lfoot{} \fancyhead[RO]{\footnotesize{Light scattering
from three-level systems}\\Tom Savels, Allard P. Mosk and Ad
Lagendijk} \cfoot{\thepage\ of \pageref{LastPage}}
\pagestyle{fancy}
%\numberwithin{equation}{section}

\section{\label{sec:intro}Introduction}
In the description of the interaction of atomic ensembles with
light, the atoms are often treated in the electric dipole
approximation, and are therefore effectively considered to be
point-dipoles. The advantages of the point-dipole formalism are
twofold. Firstly, the delta-function potential associated with
point-scatterers allows for significant mathematical
simplifications compared to finite-size scatterers \cite{P d
Vries}. Secondly, the point-scatterer formalism allows for a
transparent description of many multiple-scattering phenomena,
mimicking most of the associated relevant physics \cite{B v
Tiggelen}. The light scattering properties of point-dipoles can be
expressed by means of their T-matrix $\tensor{T}$ which is related
to the dynamic polarizability $\tensor{\alpha}$ by
$\tensor{T}(\omega,{\bm r}',{\bm
r})=-(\omega/c)^2\tensor{\alpha}(\omega)\delta({\bm r})\delta({\bm
r}-{\bm r}')$. If one describes the internal structure of the
atoms as an effective two-level system \cite{L Allen} or a damped
harmonic oscillator, one finds the well-known Lorentzian
expression for the linear dynamic polarizability \cite{R Laudon}
for frequencies near the resonance frequency.\\
\indent In this paper, we present a extension of the point-dipole
T-matrix to three-level atoms. Three-level atoms have been studied
a lot in the past, especially in the context of lasing without
inversion \cite{K Meduri, A Popov}. However, in this paper, we are
focussing on a different application of three-level atoms. Our aim
is to develop a transparent formulation of a point-dipole with
gain, based on clear physical grounds. The introduction of a pump
in the point-dipole model is highly attractive because it allows,
within the framework of the T-matrix formalism, for a description
of optically amplifying atomic systems \cite{J Kimble, K An, D
Meschede, C Ginzel}. The results presented in this paper are a
first step towards a better qualitative and quantitative
understanding of the presence of gain in multiple light scattering
systems, such as atomic lattices \cite{I Bloch} or coherent
backscattering experiments \cite{C Miniatura}.
\\
\indent Our paper is organized as follows. In Section II, the
Master equation for an optically pumped three-level system is
solved. The evolution of all atomic populations and internal
coherences can then be deduced. In particular, we will focus on
the system's steady-state regime. In Section III, we will connect
the reduced density matrix of the system with its dynamic
polarizability, which will lead us to the key result of this
paper: the T-matrix of a point-dipole with gain. In Section IV,
finally, we will discuss the physical meaning of our result, and
elaborate on the dispersion and dissipation of a dipole in the
presence of gain.
\\
\section{\label{sec:Master}Master equation of a pumped three-level
system}
\begin{figure}[t]
        \begin{center} \includegraphics[width=3in]{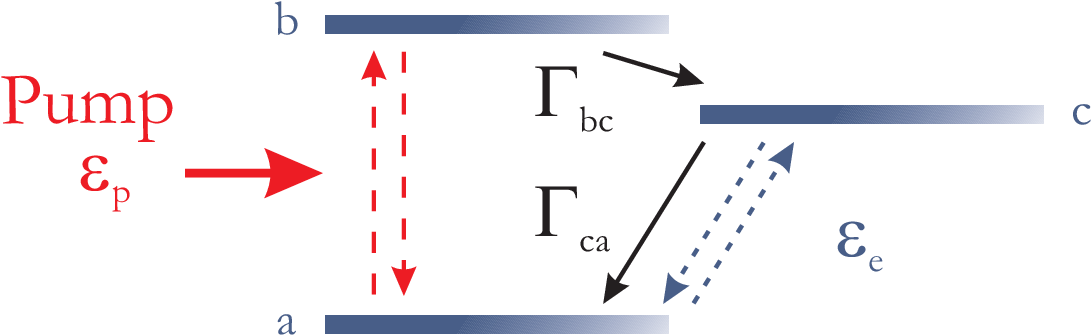}
        \caption{\label{fig,Three_Levels}(Color online) The three-level system \textit{A}.
        Decay from $b$ to $c$ and from $c$ to $a$ can occur according to the decay rates $\Gamma_{bc}$ and $\Gamma_{ca}$, respectively.
        Decay from $b$ to $a$ is neglected. The left dashed arrows express
        the interaction with the pump. The right dashed arrows depict the
        interaction with the external probe field.}
        \end{center}
\end{figure}
We start by considering a three-level system \textit{A}, shown in
Figure \ref{fig,Three_Levels}. The experimental situation we have
in mind is, e.g., a hydrogen-like finestructureless atom in a
magnetic field. The $c\rightarrow a$ transition then corresponds
to an allowed $\pi$-transition, whereas level $b$ is a higher
lying quickly decaying level. Decay between the energy levels
levels $a$, $b$ and $c$ can occur according to the decay rates
$\Gamma_{bc}$ and $\Gamma_{ca}$. The $b\rightarrow c$ transition
either has a non-radiative nature, or its transition frequency is
far detuned compared to the other relevant frequencies in the
system. In this paper, we are interested in the simplest scheme
possible that allows for incoherent pumping. Therefore, we will
make the following simplifying assumptions. We assume that
spontaneous emission from level $b$ to level $a$ can be neglected
compared to other decay processes, hence $\Gamma_{ba} \approx 0$.
Furthermore, the lifetime $\Gamma_{bc}^{-1}$ of level $b$ is
chosen to be small compared to the lifetime $\Gamma_{ca}^{-1}$.
Both previous restrictions on the decay rates do not affect the
final results of this paper in any qualitative way, and are only
introduced for mathematical simplicity. Of course, quantitative
differences will arise if extra coherences imposed by the pump are
taken into account through a finite $\Gamma_{ba}$, but these
effects are outside the scope of this paper. Two incident
electromagnetic fields ${\bm \varepsilon}_{e}$ and ${\bm
\varepsilon}_{p}$ interact with \textit{A}. The field ${\bm
\varepsilon}_{p}$ serves as a pump to establish population
inversion in the two-level system \textit{ac}. The external probe
field ${\bm \varepsilon}_{e}$ is present in order
to calculate the dynamic polarizability of the \textit{ac} system.\\
\indent We are interested in the steady-state behavior of the
system \textit{A}. We will now derive the Master equation of the
system, using the standard procedure presented in \cite{C
Cohen-Tannoudji}. The total Hamiltonian of the system \textit{A},
the electromagnetic fields and interactions can be written in the
electric dipole and the rotating wave approximation as
\\
\begin{align}
    \hat{H} \equiv
    \hat{H}_{A}+\hat{H}_{P}+\hat{H}_{R}+\hat{V}_{AR}+\hat{V}_{AE}+\hat{V}_{AP}.\label{sec,Master,H_total}
\end{align}
\\
The atomic Hamiltonian $\hat{H}_{A}$ is given by
$\hat{H}_{ab}+\hat{H}_{c}$, with
\\
\begin{subequations}
\label{sec,Master,H_A}
\begin{align}
\hat{H}_{ab}& \equiv \hbar
\omega_{1}\hat{S}_{ab}^{+}\hat{S}_{ab}^{-},\qquad \hat{H}_{c}
\equiv \hbar \omega_{2}\hat{S}_{ac}^{+}\hat{S}_{ac}^{-},
\end{align}
\end{subequations}
\\
with $\hat{S}_{ij}^{+}$, $\hat{S}_{ij}^{-}$ the $ij$ dipole
raising and lowering operators. The Hamiltonian of the pump is
given by
\\
\begin{align}
\hat{H}_{P}& \quad \equiv \quad \hbar \omega_{p}\Bigl(\hat{a}_{p}
\hat{a}^{\dagger}_{p} + \tfrac{1}{2}\Bigr),\label{sec,Master,H_P}
\end{align}
\\
where the operators $\hat{a}_{p}$ and $\hat{a}^{\dagger}_{p}$
respectively annihilate and create a pump photon with frequency
$\omega_{p}$. The system \textit{A} is coupled via the interaction
$\hat{V}_{AR}$
\\
\begin{align}
\hat{V}_{AR}& \quad \equiv -i\hat{{\bm d}}\cdot\sum_{{\bm
k}\lambda}\Bigl(\frac{\hbar \omega_{{\bm
k}\lambda}}{2\varepsilon_{0} V}\Bigr)^{1/2}\Bigl(\hat{a}_{{\bm
k}\lambda}-\hat{a}_{{\bm
k}\lambda}^{\dagger}\Bigr)\label{sec,Master,V_AR}
\end{align}
\\
to the three-dimensional multimode electromagnetic field with Hamiltonian \\
\begin{align}
\hat{H}_{R}& \quad \equiv \sum_{{\bm k}\lambda}\hbar \omega_{{\bm
k}\lambda}\Bigl(\hat{a}_{{\bm k}\lambda} \hat{a}_{{\bm
k}\lambda}^{\dagger} + \tfrac{1}{2}\Bigr).\label{sec,Master,H_R}
\end{align}
\\
The dipole operator is denoted as $\hat{\bm d}$, the operators
$\hat{a}$ and $\hat{a}^{\dagger}$ are the photon annihilation and
creation operators, and $V=L^3$ is the quantization volume. The
external field ${\bm \varepsilon}_{e}$ has a frequency $\omega$,
and interacts with the system \textit{A} by
\\
\begin{align}
\hat{V}_{AE}& \quad \equiv \quad
\frac{1}{2}\hbar\Omega_{e}\Bigl(\hat{S}_{ac}^{+}e^{-i\omega
t}+\hat{S}_{ac}^{-}e^{i\omega t}\Bigr),\label{sec,Master,V_AE}
\end{align}
\\
where the interaction strength is given by the Rabi frequency
$\Omega_{e}\equiv-{\bm d}_{ac}\cdot{\bm \varepsilon}_{e}/\hbar$,
with ${\bm d}_{ac}$ the $ac$ transition dipole moment.\\
\indent The interaction Hamiltonian $\hat{V}_{AP}$, finally,
denotes the coupling of the atom to the pump:
\\
\begin{align}
\hat{V}_{AP}& \quad \equiv \quad
g\Bigl(\hat{S}_{ab}^{+}+\hat{S}_{ab}^{-}\Bigr)\Bigl(\hat{a}_{p}+
\hat{a}^{\dagger}_{p}\Bigr),\label{sec,Master,V_AP}
\end{align}
\\
with $g$ expressing the coupling strength between the pump and the
system \cite{C Cohen-Tannoudji}.
\begin{figure}[t]
    \subfigure[The three-level system $abc$.]{
    \label{fig,Uncoupled}
    \includegraphics[width=2.5in]{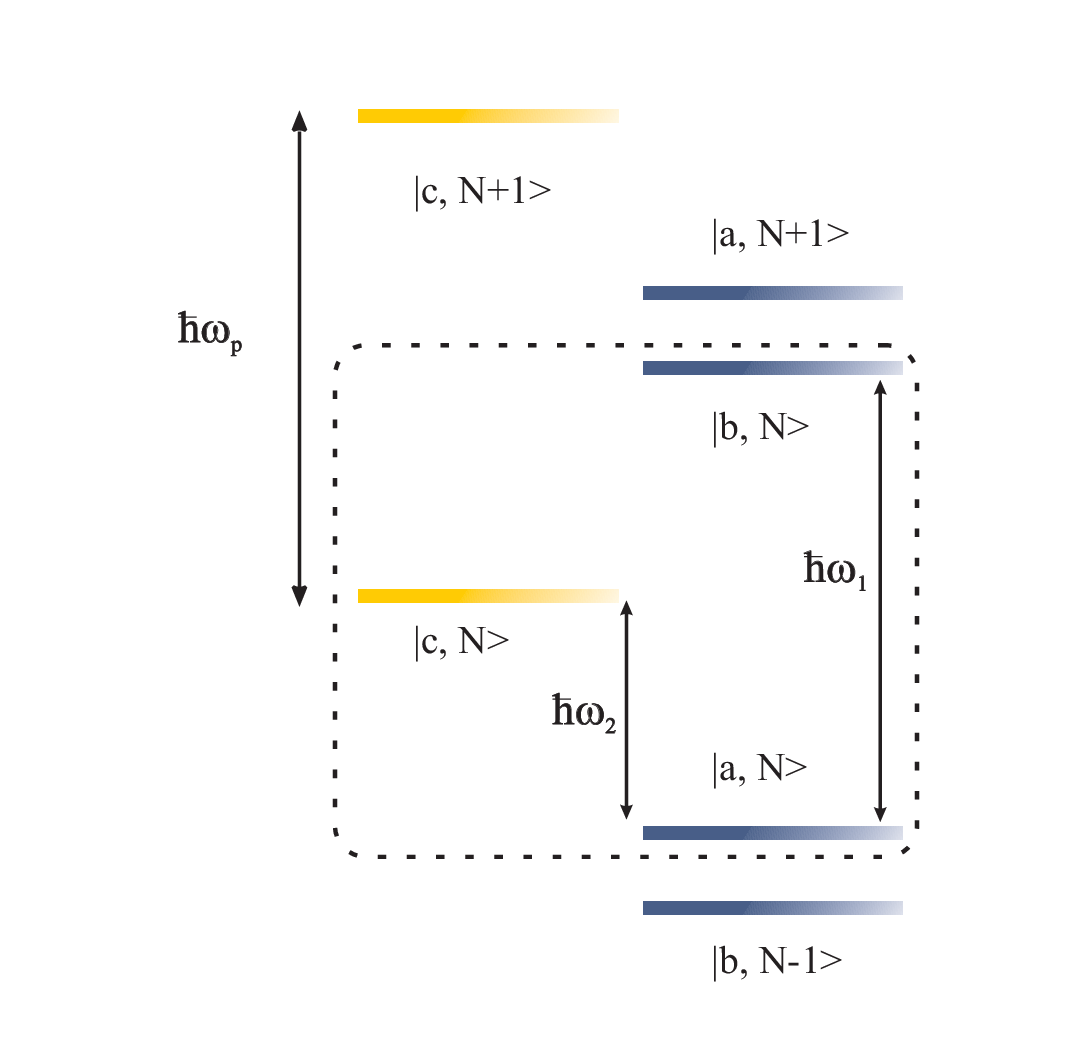}
    }
    \subfigure[The dressed states $\left|1(N)\right>$ and $\left|2(N)\right>$.]{
    \label{fig,Coupling}
    \includegraphics[width=2.5in]{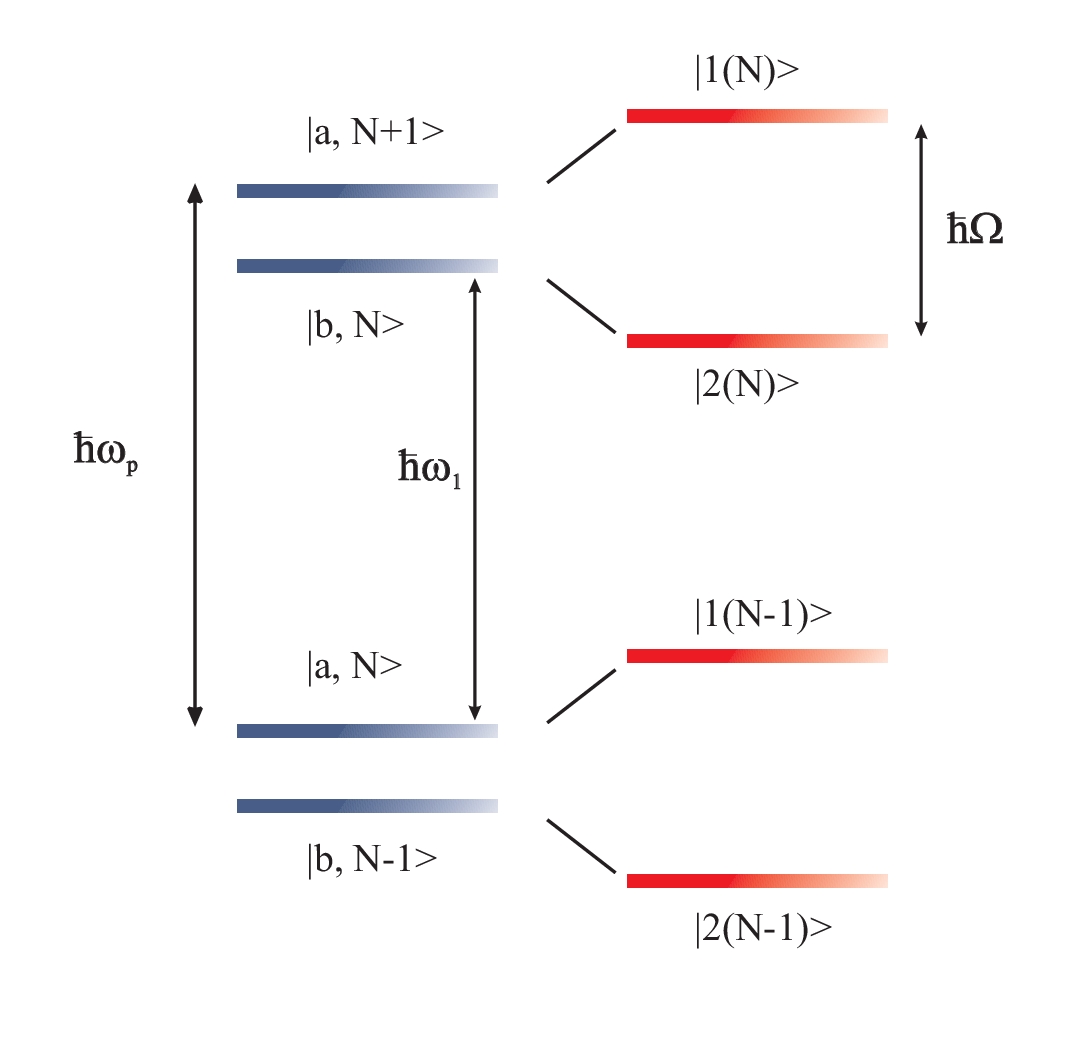}
    }
    \caption{\label{fig,Coupling_vs_uncoupling}(Color online)
    In Figure (a), levels $\left|a,N\right>$, $\left|b,N\right>$ and
    $\left|c,N\right>$ are shown for different number of pump photons $N$. The energy
    differences between energy levels are also indicated. In Figure (b), the dressed
    states $\left|1(N)\right>$ and $\left|2(N)\right>$ are schematically shown.
    The energy difference between both states is given by $\hbar\Omega=\hbar\sqrt{\delta_p^2+\Omega_{p}^2}$,
    which can easily be deduced from definitions (\ref{sec,Master,dressed_states}).}
\end{figure}
The pump field shifts the energies of the levels $a$ and $b$. This
shift, which can be significant for large pump intensities, can be
taken into account (see, e.g., \cite{H Carmichael}) by considering
the ``dressed states'' or eigenstates of the Hamiltonian
$\hat{H}_{D}\equiv\hat{H}_{ab}+\hat{H}_{P}+\hat{V}_{AP}$, given
by:
\\
\begin{subequations}
\label{sec,Master,dressed_states}
\begin{align}
\ket{1(N)} &\equiv \sin(\theta) \ket{a,N+1}+\cos(\theta) \ket{b,N},\\
\ket{2(N)} &\equiv \cos(\theta) \ket{a,N+1}-\sin(\theta)
\ket{b,N},
\end{align}
\end{subequations}
\\
graphically represented in Figure
\ref{fig,Coupling_vs_uncoupling}. Since the energy of level $c$ is
unaffected by the pump, we use the notation
$\ket{c(N)}\equiv\ket{c,N}$. The number of pump photons is then
given by N, and
\\
\begin{align}
\tan{2\theta}\equiv-\frac{\Omega_{p}}{\delta_{p}},\qquad 0\leq
2\theta < \pi,
\end{align}
\\
where we used
\\
\begin{align}
&\hbar\Omega_{p} \equiv -{\bm d}_{ab} \cdot {\bm \varepsilon}_{p},
&\delta_{p} \equiv \omega_{p}-\omega_{1},\end{align}
\\
with ${\bm d}_{ab}$ the $ab$ transition dipole moment. We assume
for now that the distribution of pump photons is relatively narrow
around a large average value ($\left<N\right>\gg\Delta N\gg 1$),
for which $2g\sqrt{\left<N\right>}=\hbar\Omega_{p}$ holds (see,
e.g., \cite{C Cohen-Tannoudji}).
\\
\indent The dynamics of the system \textit{A} can be expressed in
terms of the Master equation for the reduced density matrix
$\hat{\sigma}$. The Master equation is in the usual Born-Markov
approximation, written in the Lindblad form, given by \cite{G
Lindblad, A Rau, M Scully}
\\
\begin{align}
\frac{d}{dt}\hat{\sigma}&\equiv\hat{\mathcal{L}}\hat{\sigma}\nonumber\\
&\equiv\hat{\mathcal{L}}_{nd}\hat{\sigma}+\hat{\mathcal{L}}_{d}\hat{\sigma}\label{sec,Master,Lindblad}.
\end{align}
\\
The non-dissipative part of the Lindblad operator can be written
as
\\
\begin{align}
    \hat{\mathcal{L}}_{nd}\hat{\sigma}& \equiv -\frac{i}{\hbar}[\hat{
    H}_{D}+\hat{H}_{c}+\hat{V}_{AE},\hat{\sigma}],\label{sec,Master,Mastereq_NonDissipative}
\end{align}
\\
while the dissipative part is given by
\\
\begin{align}
    \hat{\mathcal{L}}_{d}\hat{\sigma} \equiv&
    -\frac{\Gamma_{ca}}{2}\left(\hat{S}_{ac}^{+}\hat{S}_{ac}^{-}\hat{\sigma}+\hat{\sigma}\hat{S}_{ac}^{+}\hat{S}_{ac}^{-}\right)+ \Gamma_{ca}\hat{S}_{ac}^{-}\hat{\sigma}\hat{S}_{ac}^{+}\nonumber\\
    &
    -\frac{\Gamma_{bc}}{2}\left(\hat{S}_{bc}^{+}\hat{S}_{bc}^{-}\hat{\sigma}+\hat{\sigma}\hat{S}_{bc}^{+}\hat{S}_{bc}^{-}\right) +
    \Gamma_{bc}\hat{S}_{bc}^{-}\hat{\sigma}\hat{S}_{bc}^{+}\label{sec,Master,Mastereq_Dissipative}.
\end{align}
\\
The explicit evolution equations for all reduced density matrix
elements are found by expanding equation
(\ref{sec,Master,Lindblad}) in the basis \{$\ket{1(N)}$,
$\ket{2(N')}$, $\ket{c,N''}$\}, where $N$, $N'$ and $N''$ are not
necessarily equal. However, we will show in the next section that
only an expansion in the basis \{$\ket{1(N)}$, $\ket{2(N)}$,
$\ket{c,N+1}$\} will be required to determine the system's
T-matrix, and we will therefore restrict ourselves to this
particular basis. If we introduce the following notations for
typographical simplicity
\\
\begin{subequations}
\begin{align}
\sigma_{i,j}^{N,M}\quad\equiv&\quad\bra{i(N)}\hat{\sigma}\ket{j(M)},\\
E_{i}^N\quad\equiv&\quad\bra{i(N)}\hat{H}_{D}+\hat{H}_{c}\ket{i(N)},\\
&\qquad\qquad\qquad i,j\in\{1,2,c\},\nonumber
\end{align}
\end{subequations}
\\
we can focus on the evolution equations obtained by an expansion
of (\ref{sec,Master,Lindblad}). Firstly, the evolution of the
population of the dressed state $\left|1(N)\right>$ is given by
\\
\begin{align}
\overset{.}{\sigma}_{1,1}^{N,N}&=-i\frac{\Omega_{e}}{2}\sin(\theta)\Bigl(e^{i\omega
t}\sigma_{c,1}^{N+1,N} -
 e^{-i\omega t}\sigma_{1,c}^{N,N+1}\Bigr)\nonumber\\
 & \quad +\Gamma_{ca}\sin^2(\theta)\sigma_{c,c}^{N+1,N+1}
 -\Gamma_{bc}\cos^2(\theta)\sigma_{1,1}^{N,N}\nonumber\\
 & \quad +\frac{\Gamma_{bc}}{2}\sin(\theta)\cos(\theta)\Bigl(\sigma_{1,2}^{N,N}+\sigma_{2,1}^{N,N}\Bigr),\label{sec,Master,sigma_11}
\end{align}
\\
with similar expressions for the populations
${\sigma}_{2,2}^{N,N}$ and ${\sigma}_{c,c}^{N,N}$. Secondly,
coherences such as ${\sigma}_{1,2}^{N,N}$ evolve as
\\
\begin{align}
 \overset{.}{\sigma}_{1,2}^{N,N}&= -i\frac{\Omega_{e}}{2}\Bigl(e^{i\omega t}\sin(\theta)\sigma_{c,2}^{N+1,N} -
 e^{-i\omega t}\cos(\theta)\sigma_{1,c}^{N,N+1}\Bigr)\nonumber\\
 & \quad +\Gamma_{ca}\sin(\theta)\cos(\theta)\sigma_{c,c}^{N+1,N+1}\nonumber\\
 & \quad -\frac{\Gamma_{bc}}{2}
         \Bigl(\sigma_{1,2}^{N,N}-\sin(\theta)\cos(\theta)\sigma_{2,2}^{N,N}
         \nonumber\\
 & \qquad \qquad -\sin(\theta)\cos(\theta)\sigma_{1,1}^{N,N}\Bigr) \nonumber\\
 & \quad -\frac{i}{\hbar}\Bigl(E_{1}^N-E_{2}^N\Bigr)\sigma_{1,2}^{N,N},\label{sec,Master,sigma_12}
\end{align}
\\
with analogous expressions for all other coherences, which are
derived in the Appendix. Expanding the Master equation allows us
to describe the dynamics of our system, and in particular to look
at its the steady-state behavior. If we examine the Master
equation more closely, then two types of evolutions can be
distinguished. First, the coherences are driven by the external
probe field; therefore their evolution can be separated in a
quickly oscillating character (evolving typically at $\omega$) and
a slowly decaying envelope (evolving typically at
$\Gamma_{ca}^{-1}$). Hence, a time-independent regime or
``steady-state'' for $\hat{\sigma}(t)$ can only be obtained for
the envelope of the coherences. The other matrix elements (the
populations) however, do not exhibit such quick oscillatory
behavior and decay to their steady-state value without any
persistent oscillations. The steady-state value of the level
populations $\pi_{i}$, $i\in\{a,b,c\}$, for example, is given in
the small external probe field limit by
\\
\begin{subequations}
\begin{align}
\pi_{c}^{st}& \equiv
\sum_{N}\sigma_{c,c}^{N,N}\nonumber\\
&=\frac{\Gamma_{bc}}{\Gamma_{ca}}\pi_{b}^{st}=
\frac{W}{1+W(1+2\frac{\Gamma_{ca}}{\Gamma_{bc}})}=
1-\pi_{b}^{st}-\pi_{a}^{st},
\end{align}
\end{subequations}
\\
where we defined the dimensionless pumping parameter
\\
\begin{align}
W \equiv \frac{\Omega_{p}^2}{\Gamma_{bc}\Gamma_{ca}},\qquad 0\leq
W<+\infty,
\end{align}
\\
such that population inversion occurs in the $ac$ system for
$W\geq1$. Now we can describe the dynamics of the system $A$, we
will focus on the connection between the steady-state solution of
the Master equation and the scattering properties of the system
for radiation near the $c\rightarrow a$ resonance.\\\\
\section{\label{sec:Tmatrix}Derivation of the T-matrix for a dipole with gain}
The Master equation allows us to express expectation values of
atomic operators in terms of reduced density matrix elements. In
the case of a single atom, the expectation value can be
interpreted as a time-average due to the statistical character of
the Master equation \cite{L Mandel}. The (time-)average of the
operator $\hat{{\bm d}}$ associated with the \textit{ac} dipole
transition, for example, is
\\
\begin{align}
\left<\hat{{\bm d}}\right>& \equiv \left<\hat{{\bm d}}_{-}\right>+\left<\hat{{\bm d}}_{+}\right>\nonumber\\
&= {\bm
d}_{ac}\bigl(Tr(\hat{\sigma}\hat{S}_{ac}^{-})+Tr(\hat{\sigma}\hat{S}_{ac}^{+})\bigr)\label{sec,Master,d_average},
\end{align}
\\
and
\\
\begin{align}
Tr(\hat{\sigma}\hat{S}_{ac}^{-})&  \equiv\sum_{N}\sum_{i=1,2,c}\bra{i(N)}\hat{\sigma}\hat{S}_{ac}^{-}\ket{i(N)}\nonumber\\
&=
\sum_{N}\bigl({\sigma}_{c,1}^{N,N-1}\sin(\theta)+{\sigma}_{c,2}^{N,N-1}\cos(\theta)\bigr),\label{sec,Master,dipole_operator}
\end{align}
\\
from which we can see that indeed only the reduced density matrix
elements given by an expansion of (\ref{sec,Master,Lindblad}) in
the basis \{$\ket{1(N)}$, $\ket{2(N)}$, $\ket{c,N+1}$\} appear in
the expression for the average dipole moment. Furthermore, if we
assume the pump frequency to be on resonance of the $ab$
transition, we can deduce that in steady-state (defining
$\omega_2-\omega\equiv \delta$):
\\
\begin{align}
&\left<\hat{{\bm d}}_{-}\right>=\nonumber\\\nonumber\\
&-{\bm d}_{ac}{e}^{-i\omega
t}\frac{1-W}{1+W}\frac{\Omega_{e}}{2\delta-i\Gamma_{ca}(1+W)+\frac{2\Omega_{e}^2}{2\delta+i\Gamma_{ca}(1+W)}}\label{sec,Tmatrix,T_with_sat}.
\end{align}
\\
\indent The (time-)averaged atomic $ac$ dipole operator is related
to the dynamic polarizability $\tensor{\alpha}$ of the $ac$ system
by (see, e.g., \cite{R Laudon})
\\
\begin{align}
\left<\hat{\bm
d}\right>&\equiv\varepsilon_{0}\text{Re}\Bigl[\tensor{\alpha}(\omega)\cdot{\bm
\varepsilon}_{e}e^{-i\omega t}\Bigr]\label{sec,TMatrix,d_average}.
\end{align}
\\
Both equivalent expressions (\ref{sec,Master,d_average}) and
(\ref{sec,TMatrix,d_average}) allow us to connect the
polarizability $\tensor{\alpha}$ with the reduced density matrix,
yielding
\\
\begin{align}
\tensor{\alpha}(\omega)\cdot{\bm
\varepsilon}_{e}&=-\frac{1}{\hbar\varepsilon_{0}}{\bm d}_{ac}({\bm
d}_{ac}\cdot{\bm
\varepsilon}_{e})\frac{1-W}{1+W}\times\nonumber\\
&\qquad\frac{1}{-\delta+i\frac{\Gamma_{ca}}{2}(1+W)-\frac{2\Omega_{e}^2}{2\delta+i\Gamma_{ca}(1+W)}}\label{alpha},
\end{align}
\\
which simplifies for small external probe fields to
\\
\begin{align}
\tensor{\alpha}(\omega)&=-\tensor{\alpha}_{0}\frac{1-W}{1+W}\frac{1}{2}\frac{\omega_{2}}{\omega-\omega_{2}+i\frac{\Gamma_{ca}}{2}(1+W)}.
\end{align}
\\
The static polarizability is given by
\\
\begin{align}
\tensor{\alpha}_{0}\equiv\alpha_{0}{\bm \mu}\otimes{\bm
\mu},\qquad \alpha_{0}\equiv
\frac{2}{\omega_{2}\hbar\varepsilon_{0}},
\end{align}
\\
where ``$\otimes$'' denotes the tensor product of two vectors, and
${\bm \mu}$ is the unit vector parallel to ${\bm d}_{ac}$.\\
\indent It is important to note that the same expression
(\ref{alpha}) is obtained if the dressing of the levels $a$ and
$b$ is omitted and the optical Bloch equations are used instead of
the Master equation. The incoherent pumping mechanism
only appears in the polarizability as a parameter $W$, without causing any detuning effects.\\
\indent The scattering properties of a point-dipole - or, more
generally, any scattering object - can be expressed by its
T-matrix $\tensor{T}$ (see, e.g., \cite{P d Vries}). The T-matrix
is closely connected to the dynamic polarizability of the
scatterer. For a point-dipole located at ${\bm r}=0$, both
scattering quantities are related by
\\
\begin{align}
\bra{{\bm r}}\tensor{T}(\omega)\ket{{\bm r}'} &=
-\Bigl(\frac{\omega_{2}}{c}\Bigr)^2\tensor{\alpha}(\omega)\delta({\bm r})\delta({\bm r}-{\bm r}')\nonumber\\
& = t(\omega){\bm \mu}\otimes{\bm \mu}\delta({\bm r})\delta({\bm
r}-{\bm r}'),\label{sec,TMatrix,T_Abstract}
\end{align}
\\
with as matrix element $t(\omega)$:
\\
\begin{align}
t(\omega)\equiv &\alpha_{0}\Bigl(\frac{\omega_{2}}{c}\Bigr)^2\frac{1-W}{1+W}\frac{1}{2}\times\nonumber\\
&\frac{\omega_{2}}{\omega-\omega_{2}+i\frac{\Gamma_{ca}}{2}(1+W)-\frac{\Omega_{e}^2}{2\omega_2-2\omega+i\Gamma_{ca}(1+W)}},\label{sec,TMatrix,T,intermediate}
\end{align}
\\
which is nonlinear in the incident probe field (through
$\Omega_{e}$). Both delta functions appearing in
(\ref{sec,TMatrix,T_Abstract}) express the local character of the
scatterer; the anisotropy of the T-matrix is clearly due to the
preferential orientation induced by the transition dipole moment.
In the absence of pumping ($W=0$) and for small external probe
fields, we recover the expression for the linear dynamic
polarizability of a two-level atom, which satisfies the optical
theorem \cite{B v Tiggelen} expressing energy conservation:
\\
\begin{align}
-\text{Im}\Bigl[\frac{t(\omega)}{\omega_2/c}\Bigr]\Biggr|_{\substack{W=0\\\Omega_{e}\rightarrow
0}}
&=\frac{|t(\omega)|^2}{6\pi}\Biggr|_{\substack{W=0\\\Omega_{e}\rightarrow
0}}.\label{optical_theorem}
\end{align}
\\
For the optical theorem to hold, the static polarizability must
satisfy
\\
\begin{align}
\alpha_{0}=\frac{6\pi}{(\omega_{2}/c)^3}\frac{\Gamma_{ca}}{\omega_{2}},
\end{align}
\\
which, if substituted in (\ref{sec,TMatrix,T,intermediate}),
yields the final expression for the T-matrix element of a
point-dipole with gain for arbitrary pump and external probe field
intensity. In the limit for small external probe fields,
(\ref{sec,TMatrix,T,intermediate}) then reduces to
\\
\begin{align}
t(\omega) &=
\frac{3\pi}{\omega_{2}/c}\frac{\Gamma_{ca}}{\Bigl(\omega-\omega_{2}+i\frac{\Gamma_{ca}}{2}(1+W)\Bigr)}\frac{1-W}{1+W}.\label{sec,TMatrix,T_Final}
\end{align}
\\
Expressions (\ref{sec,TMatrix,T_Abstract}) and
(\ref{sec,TMatrix,T_Final}) (or more generally,
(\ref{sec,TMatrix,T,intermediate})) are the key results of this
paper. In the next section, we will elaborate on their properties
and physical consequences.
\section{\label{sec:Tmatrix Properties}Properties of the T-matrix for a dipole with gain}
The T-matrix (\ref{sec,TMatrix,T_Abstract}) fully expresses the
scattering properties of a point-dipole with gain and satisfies
the Kramers-Kronig relations \cite{N Ashkroft}, which can be
easily verified.\\
\indent The pump influences the T-matrix in a clear physical way.
The most obvious effect of the pump is to induce changes of the
(time-)averaged populations of the levels $a$ and $c$, which leads
to the multiplication of the dynamic polarizability with a factor
$(1-W)/(1+W)$. Remarkably, the multiplication factor becomes
negative if population inversion is present in the $ac$ system
($W>1$), expressing the fact that not only the imaginary, but also
the real part of the dynamic polarizability is drastically changed
by pumping. In other words, not only does absorption change into
gain, but the dispersion relation also changes, as is shown in
Figure \ref{fig,dispersion}.
\begin{figure}
    \subfigure[$W=0$]{
    \label{fig,dispersion,normal}
    \includegraphics[width=3in]{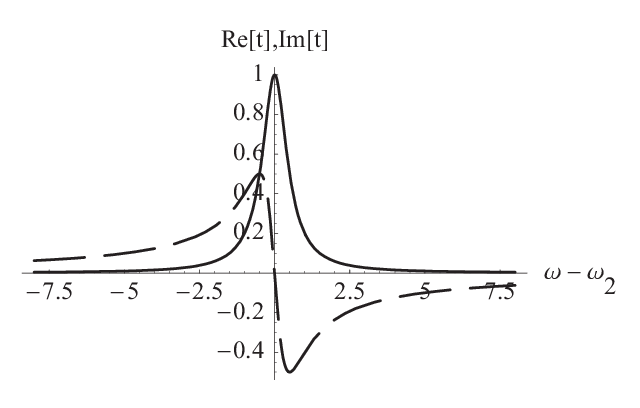}
    }
    \subfigure[$W=3$]{
    \label{fig,dispersion,flipped}
    \includegraphics[width=3in]{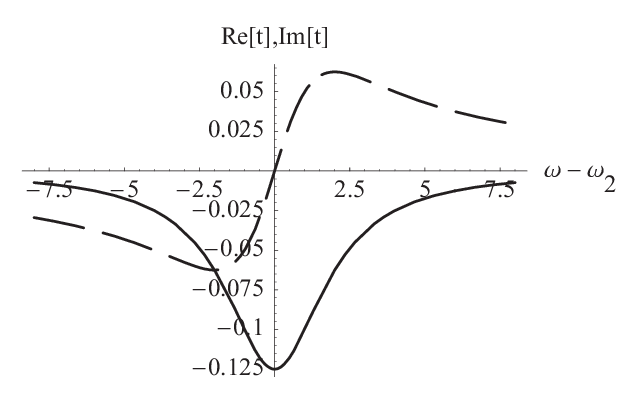}
    }
    \caption{\label{fig,dispersion}(Color online) The absorption (solid line),
    described by $\text{Im}[t(\omega)]$, and the dispersion (dashed line),
    given by $\text{Re}[t(\omega)]$ for
    the pumped three-level atom we consider, as a function of the detuning $\omega-\omega_2$
    in units of $\Gamma_{ca}$ (a) without a pump ($W=0$) (b) for a pumping
    intensity in the population inversion regime ($W=3$).
    The graphs are scaled such
    that $\text{Im}[t(\omega_2)]=1$ in Figure (a).
    }
\end{figure}
Besides changing the sign of the dynamic polarizability, the gain
also effectively broadens the $a$ state by a factor $(1+W)$, which
is equivalent with a decrease of the $Q$-factor of the
$c\rightarrow a$ resonance.
\\
\begin{figure}[t]
        \begin{center} \includegraphics[width=3in]{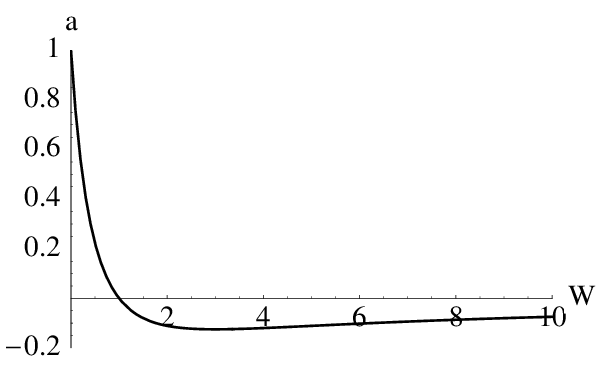}
        \caption{\label{fig,OpticalTheorem}(Color online) The albedo $a$ of the $ac$ system as a function of the reduced
        pumping parameter $W$.}
        \end{center}
\end{figure}
\indent Secondly, it is intuitively clear that the optical theorem
does not hold any longer as a nonzero pump is applied. We will now
show that this is indeed the case. If an external probe field with
polarization vector ${\bm \epsilon}$ is incident on the
point-dipole we consider here, then the dipole scattering
cross-section $\sigma_{sca}$ and extinction cross-section
$\sigma_{ext}$ are given by
\\
\begin{subequations}
\label{sec,TMatrix,cross-sections}
\begin{align}
\sigma_{sca}&=+\frac{({\bm \mu}\cdot{\bm
\epsilon})^2}{6\pi}|t(\omega)|^2,\\
\sigma_{ext}&=-\frac{({\bm \mu}\cdot{\bm
\epsilon})^2}{\omega/c}\text{Im}{\Bigl[t(\omega)\Bigr]}.
\end{align}
\end{subequations}
\\
Both scattering cross-sections depend (through $t(\omega)$) on the
applied pumping intensity. The albedo of the $ac$ system is
defined as the ratio of the elastic scattering cross-section and
the extinction cross-section, which can be written for small
incident probe fields as
\\
\begin{align}
a\equiv\frac{\sigma_{sca}}{\sigma_{ext}}=\frac{1-W}{(1+W)^2},\label{sec,TMatrix,albedo}
\end{align}
\\
where $(1-a)$ is the fraction of the incident probe field which is
taken away from the incident beam but not transformed into
scattered light. If the optical theorem is satisfied, the albedo
is one, as can be seen from expression (\ref{optical_theorem}).
Therefore, if the optical theorem does not hold any longer, we
expect $a$ to be smaller than unity.
% (obviously, the albedo can
%never be larger than unity since no more light can be scattered
%than is taken away from the incident probe field)
To show that
this is indeed true, we plotted the albedo
(\ref{sec,TMatrix,albedo}) in Figure \ref{fig,OpticalTheorem} as a
function of the reduced pumping parameter $W$. As soon as the
dipole has internal population inversion ($W>1$), the albedo
becomes negative, which indicates that the point-dipole then has a
negative extinction cross-section, as we expect. Obviously, the
scattering cross-section is - by definition - always positive and
the effect of the pump is manifested purely as a decreasing of the
scattered field for increasing pump. Furthermore, the extinction
cross-section is, in absolute value, always larger than the
scattering cross-section for nonzero pump. In other words, some of
the incident probe field is taken away from the incident beam, but
not scattered elastically. The presence of inelastic scattering is
no surprise, since the total intensity emitted by the dipole is
proportional to the population of the upper state $c$, whereas the
coherent intensity emitted by the dipole is proportional to the
amount of coherence between $a$ and $c$ (see, e.g., \cite{C
Cohen-Tannoudji}). Both intensities are only equal (all scattering
is then elastic) in the absence of a pump and in the small
external probe field limit. As soon as a pump is applied, or
for larger external probe fields \cite{T Chaneliere}, some of the light is scattered inelastically.\\
\indent Finally, we see from expressions
(\ref{sec,TMatrix,cross-sections}) that the optical theorem is
also satisfied for the nontrivial values $W=1$ and
$W\rightarrow+\infty$. The latter two pumping values correspond to
pumping intensities for which the T-matrix vanishes (therefore the
optical theorem holds), but the reason why is clearly different
for both cases: for $W=1$, populations are, on average, equally
distributed among levels $a$ and $c$, preventing the building up
of an average scattered field; for $W\rightarrow+\infty$, on the
other hand, the linewidth broadening of the $c\rightarrow a$
transition induced by the pump inhibits scattering. We stress that
the absence of light scattering for $W=1$ has to be interpreted in
a statistical sense:  the T-matrix is deduced from the
(statistical) Master equation (\ref{sec,Master,Lindblad}), and is
therefore a time-averaged property of the system. In other words,
for $W=1$, the pumped point-dipole does not scatter light
\textit{on average}. Furthermore, the transparency in our system
is caused by a fully incoherent pumping scheme, contrary to, e.g.,
electromagnetically
induced transparency \cite{S Harris}. \\
\indent Finally, we note that the strong dependence of the albedo
on the applied pumping intensity could lead to interesting
experimental work on, e.g., coherent backscattering \cite{M Havey,
M Mark}, since the width of the coherent backscattering cone is
closely related to the albedo of the scatterers in the multiple
scattering medium considered.
\section{\label{sec:Summary}Summary} The aim of this paper was
to find the light scattering properties of a pumped point-dipole,
modelled as a three-level system. The resulting T-matrix
(\ref{sec,TMatrix,T_Final}) (or more generally,
(\ref{sec,TMatrix,T,intermediate})) is surprisingly intuitive. The
influence of the pump can be characterized by a single
dimensionless parameter $W$, expressing both a decrease of the
$Q$-factor and a decrease of the T-matrix. Physically, the
presence of gain not only causes the dipole to scatter partially
inelastically, but also induces important changes in its
dispersion and dissipation.
\begin{acknowledgements}
We would like to thank M. Wubs and W.L. Vos for inspiring
discussions. This work is part of the research program of the
`Stichting voor Fundamenteel Onderzoek der Materie' (FOM), which
is financially supported by the `Nederlandse Organisatie voor
Wetenschappelijk Onderzoek' (NWO).\\
\end{acknowledgements}
\appendix
\section{Expansion of the Master equation}
The aim of this appendix is to show the expansion of the Master
equation (\ref{sec,Master,Lindblad}) in the basis \{$\ket{1(N)}$,
$\ket{2(N)}$, $\ket{c,N+1}$\}. The reduced density matrix elements
which are diagonal in the atomic states evolve as
\\
\begin{align}
 \overset{.}{\sigma}_{1,1}^{N,N}&=-i\frac{\Omega_{e}}{2}\sin(\theta)\left(\sigma_{c,1}^{N+1,N}e^{i\omega t} -
 \sigma_{1,c}^{N,N+1}e^{-i\omega t}\right)\nonumber\\
 &\quad +\Gamma_{ca}\sin^2(\theta)\sigma_{c,c}^{N+1,N+1}-\Gamma_{bc}\cos^2(\theta)\sigma_{1,1}^{N,N}\nonumber\\
 & \quad
 +\frac{\Gamma_{bc}}{2}\sin(\theta)\cos(\theta)\left(\sigma_{1,2}^{N,N}+\sigma_{1,2}^{N,N}\right),\\
\nonumber\\
 \overset{.}{\sigma}_{2,2}^{N,N}&= -i\frac{\Omega_{e}}{2}\cos(\theta)\left(\sigma_{c,2}^{N+1,N}e^{i\omega t} -
 \sigma_{2,c}^{N,N+1}e^{-i\omega t}\right)\nonumber\\
 &\quad
 +\Gamma_{ca}\cos^2(\theta)\sigma_{c,c}^{N+1,N+1}-\Gamma_{bc}\sin^2(\theta)\sigma_{2,2}^{N,N}\nonumber\\
& \quad
+\frac{\Gamma_{bc}}{2}\sin(\theta)\cos(\theta)\left(\sigma_{1,2}^{N,N}+\sigma_{1,2}^{N,N}\right),\\
\nonumber\\
 \overset{.}{\sigma}_{c,c}^{N+1,N+1}&= -i\frac{\Omega_{e}}{2}\sin(\theta)\left(\sigma_{1,c}^{N,N+1}e^{-i\omega t} -
    \sigma_{c,1}^{N+1,N}e^{i\omega t}\right)\nonumber\\
    &\quad -i\frac{\Omega_{e}}{2}\cos(\theta)\left(\sigma_{2,c}^{N,N+1}e^{-i\omega t} -
 \sigma_{c,2}^{N+1,N}e^{i\omega t}\right)\nonumber\\
 &\quad -\Gamma_{ca}\sigma_{c,c}^{N+1,N+1}
 + \Gamma_{bc}\cos^2(\theta) \sigma_{1,1}^{N+1,N+1}\nonumber\\
 & \quad
 -\Gamma_{bc}\sin(\theta)\cos(\theta)(\sigma_{1,2}^{N+1,N+1}+\sigma_{2,1}^{N+1,N+1})\nonumber\\
 &\quad +\Gamma_{bc}\sin^2(\theta) \sigma_{2,2}^{N+1,N+1},
\end{align}
\\
while the elements which are off-diagonal in the atomic states
evolve as
\begin{widetext}
\begin{subequations}
\begin{align}
 \overset{.}{\sigma}_{1,2}^{N,N}&= -i\frac{\Omega_{e}}{2}\left(\sin(\theta)\sigma_{c,2}^{N+1,N}e^{i\omega t} -
 \cos(\theta){\sigma}_{1,c}^{N,N+1}e^{-i\omega t}\right)+\Gamma_{ca}\sin(\theta)\cos(\theta)\sigma_{c,c}^{N+1,N+1}\nonumber\\
&\quad -\frac{\Gamma_{bc}}{2}
(\sigma_{1,2}^{N,N}-\sin(\theta)\cos(\theta)\sigma_{2,2}^{N,N}
-\sin(\theta)\cos(\theta)\sigma_{1,1}^{N,N})-\frac{i}{\hbar}(E_{1}^N-E_{2}^N)\sigma_{1,2}^{N,N},
\\
\nonumber\\
 \overset{.}{\sigma}_{1,c}^{N,N+1}e^{-i\omega t}&= -i\frac{\Omega_{e}}{2}\left(\sin(\theta)\sigma_{c,c}^{N+1,N+1} -
 \sin(\theta){\sigma}_{1,1}^{N,N}-\cos(\theta)\sigma_{1,2}^{N,N}\right)-\frac{\Gamma_{ca}}{2}\sigma_{1,c}^{N,N+1}\nonumber\\
 & \quad -\frac{\Gamma_{bc}}{2}\left(\cos^2(\theta)\sigma_{1,c}^{N,N+1}e^{-i\omega t}-\sin(\theta)\cos(\theta)\sigma_{2,c}^{N,N+1}e^{-i\omega t}\right)+\frac{i}{\hbar}(E_{c}^{N+1}-E_{1}^{N})\sigma_{1,c}^{N,N+1}e^{-i\omega
 t},
\\
\nonumber\\
 \overset{.}{\sigma}_{2,c}^{N,N+1}e^{-i\omega t}&= -i\frac{\Omega_{e}}{2}\left(\cos(\theta)\sigma_{c,c}^{N+1,N+1} -
 \cos(\theta)\sigma_{2,2}^{N,N}-\sin(\theta)\sigma_{1,2}^{N,N}\right)-\frac{\Gamma_{ca}}{2}\sigma_{2,c}^{N,N+1}\nonumber\\
 & \quad +\frac{\Gamma_{bc}}{2}\left(\sin(\theta)\cos(\theta)\sigma_{1,c}^{N,N+1}e^{-i\omega t}-\sin^2(\theta)\sigma_{2,c}^{N,N+1}e^{-i\omega t}\right)+\frac{i}{\hbar}(E_{c}^{N+1}-E_{2}^{N})\sigma_{2,c}^{N,N+1}e^{-i\omega
 t},
\end{align}
\end{subequations}
\end{widetext}
and
\\
\begin{subequations}
\begin{align}
\overset{.}{\sigma}_{2,1}^{N,N}&=(\overset{.}{\sigma}_{1,2}^{N,N})^*\\
\overset{.}{\sigma}_{c,1}^{N+1,N}&=(\overset{.}{\sigma}_{1,c}^{N,N+1})^*\\
\overset{.}{\sigma}_{c,2}^{N+1,N}&=(\overset{.}{\sigma}_{2,c}^{N,N+1})^*.
\end{align}
\end{subequations}
\\

\end{document}